\documentclass[12pt]{article}
\usepackage{latexsym,amssymb}
\newcommand{\be}{\begin{equation}}
\newcommand{\ee}{\end{equation}}
\newcommand{\beqs}{\begin{eqnarray}}
\newcommand{\eeqs}{\end{eqnarray}}

\def\({\left(}
\def\){\right)}

\newcommand{\Exc}[1]{{${\rm E}_{{#1}({#1})}$}}

\def\mxth{\mathsurround=0pt }
\def\xversim#1#2{\lower2.pt\vbox{\baselineskip0pt \lineskip-.5pt
x  \ialign{$\mxth#1\hfil##\hfil$\crcr#2\crcr\sim\crcr}}}

\renewcommand{\a}{\alpha}
\renewcommand{\b}{\beta}

\renewcommand{\d}{\delta}

\newcommand{\g}{\gamma}

\newcommand{\m}{\mu}
\newcommand{\n}{\nu}

\def\be{\begin{equation}}
\def\ee{\end{equation}}
\def\bea{\begin{eqnarray}}
\def\eea{\end{eqnarray}}

\newcommand{\ft}[2]{{\textstyle\frac{#1}{#2}}}

\newcommand{\eqn}[1]{(\ref{#1})}
\def\bfone{\relax{\rm 1\kern-.35em 1}}

\begin{document}
\begin{titlepage}
\begin{flushright}
ITP-UU-03/53 \\
SPIN-03/34\\[1mm]
{\tt hep-th/0311224}
\end{flushright}
\vskip 16mm
\begin{center}
{\Large {\bf Maximal Supergravity from
}}\\[3mm]
{\Large {\bf IIB Flux Compactifications }}

\vskip 12mm

{\bf Bernard de Wit$^1$, Henning Samtleben$^1$ and Mario Trigiante$^2$}

\vskip 4mm

{$^1$ \em Institute for Theoretical Physics \& Spinoza Institute}\\
{\em Utrecht University, Utrecht, The Netherlands}\\[1mm]

{\tt B.deWit@phys.uu.nl}, {\tt H.Samtleben@phys.uu.nl}\\[2mm]

{$^2$ \em Dipartimento di Fisica}\\ 
{\em Politecnico di Torino, Torino, Italy} \\[1mm]
{\tt M.Trigiante@phys.uu.nl}

\vskip 6mm

\end{center}

\vskip .2in

\begin{center} {\bf Abstract } \end{center}
\begin{quotation}\noindent
Using a recently proposed group-theoretical approach, we explore novel 
gaugings of maximal supergravity in four dimensions with gauge group
embeddings that can be generated by fluxes of IIB string theory. The
corresponding potentials are positive without stationary points. Some
allow domain wall solutions which can be elevated to ten dimensions.
Appropriate truncations describe 
type-IIB flux compactifications on $T^6$ orientifolds leading to
non-maximal, four-dimensional, supergravities.  
\end{quotation}
\end{titlepage}

\eject
M/string-theory compactifications with nontrivial fluxes, possibly on
orientifolds and/or with (space-filling) D3-branes, lead to effective
field theories that take the form of gauged supergravity theories (see
e.g. \cite{G1,G2,Ga,Gx,G3,Gb,G4,G5,G6,Gy,G7,AFT}). Because of the nature of
the corresponding gauge groups not many of these gaugings have been
explored, especially not in the case of higher-extended
supergravity. The gaugings give rise to a potential for the scalar
moduli, which may freeze (some of) these moduli and may lead to broken
supersymmetry. In this paper we will be considering gaugings of
four-dimensional maximal supergravity whose gauge group can be
generated by type-IIB fluxes. Gaugings of this type are new and in
order to analyze them we make use of the group-theoretical approach
developed in \cite{dWST1,dWST3}. We do not expect that gaugings of
maximal supergravity theories are themselves of immediate
phenomenological relevance, but, upon orientifold projections, it is
possible to obtain a variety of non-maximal theories which may have
phenomenological applications.

\paragraph{Maximal gauged supergravity}
In maximal supergravity theories the effect of a gauging is encoded in
the covariant derivatives and field strengths for the kinetic terms,
and in the so-called $T$-tensor \cite{dWN82} for the potential and the
mass terms. This $T$-tensor is directly proportional to the embedding
tensor, which defines how the gauge group is embedded into the
invariance group of the theory, and plays the same role as the
superpotential and/or the moment maps that appear in non-maximal
supergravity. In four spacetime dimensions the situation is special
because of electric-magnetic duality. While the equations of motion
and the Bianchi identities are invariant under \Exc7, the Lagrangian
is only invariant under a subgroup, which we call the `electric'
subgroup. The possible gauge group has to be embedded into this
subgroup. However, the Lagrangian is not unique and neither is its
electric subgroup and this phenomenon forms somewhat of an impediment
for analyzing general gaugings. In \cite{dWST1} we have discussed this
in detail, and we have chosen to first select a Lagrangian with a
corresponding electric invariance group, prior to considering the
possible gauge group. Here we intend to show that one can also proceed
differently and start by selecting a possible $T$-tensor, without
specifying the Lagrangian. To be consistent this $T$-tensor must
belong to a particular irreducible representation of \Exc7, and its
square must take its values in a restricted number of representations.

The $T$-tensor is linearly related to the embedding tensor, which
defines the composition of the gauge group generators in terms of the
\Exc7 generators $t_\a$, where $\a=1,\ldots,133$,
\be
X_M = \Theta_M{}^\a\,t_\a\,.
\ee 
Here $X_M$ are the gauge group generators and the index
$M=1,\ldots,56$ labels the electric and magnetic charges associated
with the 28 gauge fields and their magnetic duals. The gauge group
dimension should at most be equal to 28, so that the embedding tensor,
$\Theta_M{}^\a$, acts as a corresponding projection
operator. Obviously this tensor belongs to the representation, 
\be
{\bf 56}\times {\bf 133}\to {\bf 56}+{\bf 912}+{\bf 6480}\;,
\ee
and, to have a consistent gauging, it should be restricted to
the ${\bf 912}$ representation. Furthermore, $\Theta_M{}^\a$ 
should be such that the gauge group generators form a 
subalgebra. This implies a second constraint, which in the
four-dimensional context takes the form
\be
\label{quadratic-constraint} 
\Theta_M{}^\a \,\Theta_N{}^\b \; \Omega^{MN}=0\;,
\ee 
where $\Omega^{MN}$ is the \Exc7 invariant symplectic matrix. This
constraint (which implies that the ${\bf 133}+ {\bf 8645}$
representation in the square of the embedding tensor vanishes) ensures 
that the gauge group can be embedded, upon a suitable 
electric-magnetic duality transformation, into an electric subgroup of
\Exc7 (this will be discussed in more detail in the forthcoming paper;
c.f. \cite{dWST3}). Both these constraints are \Exc7 covariant which
implies that they can be formulated in terms of the
$T$-tensor. The latter is the field-dependent tensor defined  by
\be
\label{T-theta}
T_M{}^\a[\Theta,\phi]\,t_\a  = {\cal V}_{\;M}^{-1\,N}\,\Theta_N{}^\a\,
({\cal V}^{-1}t_\a  {\cal V} )\;,
\ee
where ${\cal V}$ denotes the coset representative of ${\rm
E}_{7(7)}/{\rm SU}(8)$. When treating the embedding tensor as a
spurionic object that transforms under \Exc7, the Lagrangian and
transformation rules are formally \Exc7 invariant. Of course, when
freezing $\Theta_M{}^a$ to a constant, the invariance is broken.

\paragraph{Type-IIB Fluxes} 
We consider gaugings of $N=8$ supergravity that can in principle
be generated by the three- and five-form fluxes of the type-IIB
theory. The proper setting to discuss this is by decomposing the \Exc7
group according to ${\rm SL}(2)\times {\rm SL}(6)$. There are two
different embeddings, proceeding through two different maximal subgroups,
\bea
{\rm E}_{7(7)} \longrightarrow \cases{ {\rm SL}(8) \longrightarrow {\rm
SL}(6)\times{\rm SL}(2)\times {\rm SO}(1,1) 
\cr {\rm SL}(6)\times {\rm SL}(3) \longrightarrow {\rm SL}(6)\times{\rm
SL}(2)\times {\rm SO}(1,1)  \cr} 
\eea
The fact that these embeddings are different is confirmed by the
inequivalent branchings of the ${\bf 56}$ representation of \Exc7, 
\bea
\label{sl(6/2)-decomposition}
&&{\bf 56}\longrightarrow 
\cases{{\bf 28} + \overline{\bf 28} \longrightarrow \cr
\quad ({\bf 1}, {\bf 1})_{-3} + ({\bf 6}, {\bf 2})_{-1} +
({\bf 15},{\bf 1})_{+1} 
 + (\overline{\bf 15}, {\bf 1})_{-1}  +
(\overline{\bf 6}, {\bf 2})_{+1} +({\bf 1},{\bf 1})_{+3}   \cr
\noalign{\vskip2mm}
(\overline{\bf 6}, {\bf 3}) + ({\bf 20}, {\bf 1}) +
({\bf 6}, \overline{\bf 3})  \longrightarrow \cr
\quad (\overline{\bf 6}, {\bf 1})_{-2} + ({\bf 6}, {\bf 2})_{-1} 
 + ({\bf 20}, {\bf 1})_0 +
(\overline{\bf 6}, {\bf 2})_{+1} +({\bf 6},{\bf 1})_{+2}  
 \cr} \quad{~}
\eea
Electric and magnetic charges transform according to this ${\bf
56}$ representation. The first embedding is relevant to the IIA
theory. This is obvious from the fact that it is possible to extend 
${\rm SL}(6)$ to ${\rm SL}(7)$, which is the symmetry of the
M-theory seven-torus. Upon dimensional reduction on $T^6$, the
correspondence with the resulting gauge fields is as follows. The RR
vector field corresponds to the $({\bf 1},{\bf 1})_{+3}$
representation, the magnetic dual of the graviphotons and the
NS two-rank tensor field constitute the $({\bf 6},{\bf 2})_{-1}$,
and the RR three-rank antisymmetric tensor gives rise to the 
$({\bf 15},{\bf 1})_{+1}$ representation listed in 
\eqn{sl(6/2)-decomposition}. 

The second embedding is relevant to the IIB theory. The S-duality
group coincides with ${\rm SL}(2)$. The five-rank
tensor generates gauge fields which, together with their magnetic
duals, constitute the
$({\bf 20}, {\bf 1})_0$, the graviphotons constitute the
$(\overline{\bf 6},{\bf 1})_{-2}$ and the antisymmetric tensors the
$({\bf 6}, {\bf 2})_{-1}$ representation. 
Henceforth we restrict ourselves to this embedding and
investigate the group-theoretical structure of the embedding
tensor. We have already presented the branching of the ${\bf 56}$
representation in \eqn{sl(6/2)-decomposition}; the branching of
both the ${\bf 133}$ and the ${\bf 912}$ representations, with respect
to ${\rm SL}(6)\times{\rm SL}(2)\times {\rm SO}(1,1)$ is given in
table~\ref{912-decomposition}. 

\begin{table}
\begin{center}
{\scriptsize 
\begin{tabular}{c|ccccc} \hline
&$\!(\overline{\bf 6},{\bf 1})_{-2}\!$ & $\!({\bf 6},{\bf 2})_{-1}\!$ & 
$\!({\bf 20},{\bf 1})_{0}\!$
&$\!(\overline{\bf 6},{\bf 2})_{+1}\!$ & $\!({\bf 6},{\bf 1})_{+2}\!$
\\[.1mm]
\hline 
$({\bf 1},{\bf2})_{-3}$ 
& {~}
& $({\bf 6},{\bf 1})_{-4}$ 
& $({\bf 20},{\bf 2})_{-3}$ 
& $(\overline{\bf 6},{\bf 3}+{\bf 1})_{-2}$ 
& $({\bf 6},{\bf 2})_{-1}$
\\[1mm]
$({\bf 15},{\bf 1})_{-2}$ 
& $({\bf 6},{\bf 1})_{-4}$ 
& $({\bf 20},{\bf 2})_{-3}$ 
& $(\overline{\bf 6}\!+\!\overline{\bf 84},{\bf 1})_{-2}$ 
& $({\bf 6}\!+\!{\bf 84},{\bf 2})_{-1}$ 
& $({\bf 70}\!+\!{\bf 20},{\bf 1})_{0}$ 
\\[1mm]
$(\overline{\bf 15},{\bf 2})_{-1}$ 
& $({\bf 20},{\bf 2})_{-3}$ 
& $(\overline{\bf 6}\!+\!\overline{\bf 84},{\bf 1})_{-2}+ (\overline{\bf
6},{\bf 3})_{-2}$  
& $({\bf 6}\!+\!{\bf 84},{\bf 2})_{-1}$ 
& $({\bf 20},{\bf 3}\!+\!{\bf 1})_{0}
   +(\overline{\bf 70},{\bf 1})_0$ 
& $(\overline{\bf 6}\!+\! \overline{\bf 84},{\bf 2})_{+1}$
\\[1mm]
$({\bf 1},{\bf 1})_{0}$ 
&$(\overline{\bf 6},{\bf 1})_{-2}$ 
& $({\bf 6},{\bf 2})_{-1}$ 
& $({\bf 20},{\bf 1})_{0}$
& $(\overline{\bf 6},{\bf 2})_{+1}$ 
& $({\bf 6},{\bf 1})_{+2}$
\\[1mm]
$({\bf 35},{\bf 1})_{0}$ 
& $(\overline{\bf 6}\!+\! \overline{\bf 84},{\bf 1})_{-2}$
& $({\bf 6}\!+\! {\bf 84},{\bf 2})_{-1}$
& $({\bf 70}\!+\!\overline{\bf 70}\!+\!{\bf 20},{\bf 1})_{0}$ 
& $(\overline{\bf 6}\!+\! \overline{\bf 84},{\bf 2})_{+1}$
& $({\bf 6}\!+\! {\bf 84},{\bf 1})_{+2}$
\\[1mm]
$({\bf 1},{\bf 3})_{0}$ 
& $(\overline{\bf 6},{\bf 3})_{-2}$ 
& $({\bf 6},{\bf 2})_{-1}$
& $({\bf 20},{\bf 3})_{0}$
& $(\overline{\bf 6},{\bf 2})_{+1}$ 
& $({\bf 6},{\bf 3})_{+2}$ 
\\[1mm]
$({\bf 15},{\bf 2})_{+1}$ 
& $({\bf 6}\!+\!{\bf 84},{\bf 2})_{-1}$
& $({\bf 20},{\bf 3}\!+\!{\bf 1})_{0}+({\bf 70},{\bf 1})_0$ 
& $(\overline{\bf 6}\!+\!\overline{\bf 84},{\bf 2})_{+1}$ 
& $({\bf 6}\!+\!{\bf 84},{\bf 1})_{+2} + ({\bf 6},{\bf 3})_{+2}$ 
& $({\bf 20},{\bf 2})_{+3}$ 
\\[1mm]
$(\overline{\bf 15},{\bf 1})_{+2}$ 
& $(\overline{\bf 70}\!+\!{\bf 20},{\bf 1})_{0}$
& $(\overline{\bf 6}\!+\! \overline{\bf 84},{\bf 2})_{+1}$ 
& $({\bf 6}\!+\!{\bf 84} ,{\bf 1})_{+2}$  
& $({\bf 20},{\bf 2})_{+3}$ &
$(\overline{\bf 6},{\bf 1})_{+4}$ 
\\[1mm]
$({\bf 1},{\bf 2})_{+3}$ 
& $(\overline{\bf 6},{\bf 2})_{+1}$ 
& $({\bf 6},{\bf 3}\!+\!{\bf 1})_{+2}$ 
& $({\bf 20},{\bf 2})_{+3}$ 
& $(\overline{\bf 6},{\bf 1})_{+4}$ 
& {~}
\\ \hline
\end{tabular}
}
\end{center}
\caption{\small Possible representations of the embedding tensor
$\Theta_M{}^\a$. The first column lists the branching of the adjoint
${\bf 133}$ representation of \Exc7 (defined with upper indices) and
the upper row the branching of the ${\bf 56}$ representation (see
\eqn{sl(6/2)-decomposition}).  Only those products appear that belong
to the ${\bf 912}$ representation. The notation $({\bf m},{\bf n})_p$
indicates that we are dealing with a product representation, where
${\bf m}$ denotes the ${\rm SL}(6)$ representation, ${\bf n}$ the
${\rm SL}(2)$ representation, and $p$ the weight under ${\rm
SO}(1,1)$. The $({\bf 6},{\bf 2})_{-1}$ and $(\overline{\bf 6},{\bf
2})_{+1}$ representations appear with multiplicity 2, unlike all other
representations which appear with unit multiplicity.
}\label{912-decomposition}
\end{table}

Consider type-II fluxes on $T^6$ that are related to the five- and 
three-rank field strengths, $F_{\Lambda\Sigma\Gamma\Delta\Omega}$ (and
its related dual $F_{\Lambda\m\n\rho\sigma}$) and
$G_{\Lambda\Sigma\Gamma}{}^\tau$, where $\Lambda,\Sigma,\ldots=
1,\ldots,6$ denote the $T^6$ indices and $\tau=1,2$ denotes ${\rm
SL}(2)$ indices. The $T$-tensors correspond directly to these field
strengths, modulo multiplicative modifications by the scalar
moduli. One can derive that constant five- or three-form flux
contributions give rise to $T$-tensor components transforming in the
$({\bf 6},{\bf 1})_{-4}$ and $({\bf 20},{\bf 1})_{-3}$
representations. This implies that the embedding tensor has components
transforming in the $(\overline{\bf 6},{\bf 1})_{+4}$ and 
$({\bf 20},{\bf 1})_{+3}$ representations.\footnote{
   The arguments leading to these assignments are somewhat
   subtle and the reader may wish to consult
   \cite{dWST1,dWST3}. In principle, there are also one-form field
   strengths associated with the dilaton and the RR scalar, which
   yield an embedding tensor component in the 
   $({\bf 6}, {\bf 3})_{+2}$ representation. For simplicity, this
   possible contribution will be disregarded in the subsequent
   analysis.} 
Indeed, comparison with table~\ref{912-decomposition} reveals that the
embedding tensor contains these two representations. In the
following we will thus assume that the embedding tensor consists of
two components, belonging to these particular representations. 
Type-IIB flux compactifications are contained in this class of
embedding tensors. However, the particular origin of the
gauging does not play a crucial role in what follows. 

As explained above, the embedding tensor defines the gauge group in
the four-dimensional theory. Since we know how the ${\bf 133}$
representation decomposes, we can now indicate the structure of the
gauge group. The gauge fields are contained in the $(\overline{\bf
6},{\bf 1})_{-2}+({\bf 6},{\bf 2})_{-1}+({\bf 20},{\bf 1})_{0}$
representation and thus couple to electric charges in the conjugate
representation $({\bf 6},{\bf 1})_{+2}+(\overline{\bf 6},{\bf
2})_{+1}+({\bf 20},{\bf 1})_{0}$. Denoting the $({\bf 20},{\bf
2})_{+3}$ and the $(\overline{\bf 6},{\bf 1})_{+4}$ components of the
embedding tensor by $\theta_{\Lambda\Sigma\Gamma}{}^\tau$ and
$\theta^\Lambda$, we can decompose the gauge group generators into the
\Exc7 generators belonging to the $(\overline{\bf 15},{\bf
2})_{-1}+({\bf 15},{\bf 1})_{-2}+({\bf 1},{\bf 2})_{-3}$
representation, which we denote by $t^{\Lambda\Sigma\tau}$,
$t_{\Lambda\Sigma}$ and $t^\tau$, respectively. These generators have the
following non-vanishing commutation relations, associated with \Exc7, 
\bea
\label{E7-positive}
{[}t^{\Lambda\Sigma \tau},\,t^{\Gamma\Omega \sigma}]&=& \ft12
\,\varepsilon^{\tau\sigma}\, 
\varepsilon^{\Lambda\Sigma\Gamma\Omega\Pi\Delta} \,t_{\Pi\Delta}\;,
\nonumber \\
{[}t_{\Lambda\Sigma},\,t^{\Gamma\Omega \tau}]&=& 2\,
\d_{\Lambda\Sigma}^{\Gamma\Omega} \,t^\tau\;.
\eea
Consequently, once we fix the generators $t^{\Lambda\Sigma\tau}$, 
the generators $t_{\Lambda\Sigma}$ and $t^{\tau}$ will
follow.\footnote{
This will be demonstrated in due course. Here we note that our
normalizations are such that 
${\rm tr}[ t^{\Lambda\Sigma \tau}\,t_{\Gamma\Omega \sigma}] = -24\,
\d^\tau_\sigma 
\,\d^{\Lambda\Sigma}_{\Gamma\Omega}$, 
${\rm tr}[ t^{\Lambda\Sigma}\,t_{\Gamma\Omega}] =
-24\,\d^{\Lambda\Sigma}_{\Gamma\Omega}$ and ${\rm tr}[t_\tau
\,t^\sigma]= -12\, \delta_\tau^\sigma$. 
}  
Here it is relevant to recall that the decomposition of the ${\bf 133}$
representation in 
table~\ref{912-decomposition} refers to the conjugate representation
with upper index $\a$, whereas \Exc7 generators $t_\a$ carry a lower
index. 

The gauge algebra decomposes as follows. The generators
$X^\Lambda$ and $X_{\Lambda \tau}$ vanish and the remaining generators
$X_{\Lambda\Sigma\Gamma}$, $X^{\Lambda \tau}$ and $X_{\Lambda}$ can be
decomposed as,  
\bea
\label{2-theta} 
X_{\Lambda\Sigma\Gamma} &=& 2\, \varepsilon_{\tau\sigma}\,
\theta_{\Lambda\Sigma\Gamma}{}^\tau\, t^\sigma\;, 
\nonumber\\ 
X^{\Lambda\, \tau} &=& \ft16 \,
\varepsilon^{\Lambda\Sigma\Gamma\Omega\Pi\Delta} \, 
\theta_{\Sigma\Gamma\Omega}{}^\tau \,t_{\Pi\Delta}  + 
 \,\theta^\Lambda\, t^\tau  \;, 
\nonumber\\
X_\Lambda &=& \varepsilon_{\tau\sigma}\,
\theta_{\Lambda\Sigma\Gamma}{}^\tau  \,
t^{\Sigma\Gamma \sigma}  +   \theta^{\Sigma}\,t_{\Lambda\Sigma}\; .
\eea
Observe that there is a certain degeneracy in these definitions; not
all generators $X_M$ are linearly independent, and there are at most
20 independent generators. Eqn. \eqn{2-theta} defines the
embedding matrix 
$\Theta_M{}^\a$ in terms of $\theta_{\Lambda\Sigma\Gamma}{}^\tau$ and
$\theta^\Lambda$. The various numerical coefficients have been
determined by requiring that $\Theta_M{}^\a$ is an element of the ${\bf
912}$ representation. We will comment on this shortly. 

For the construction of the corresponding $T$-tensor one needs the
${\rm E}_{7(7)}/{\rm SU}(8)$ coset representative ${\cal V}$ in a
convenient decomposition. Let us first consider the vector $V_M =
(V^\Lambda, V_{\Lambda\tau}, V_{\Lambda\Sigma\Gamma}, V^{\Lambda\tau},
V_\Lambda)$, transforming in the ${\bf 56}$ representation and
decomposed as in the top row of table~\ref{912-decomposition}, so that
the generator $t_0$ of ${\rm SO}(1,1)$ is diagonal with eigenvalues
$\pm 2$, $\pm 1$ and $0$. The \Exc7 invariant symplectic product reads 
\be
\Omega^{MN}\,V_M\,W_N = V^\Lambda\,W_\Lambda - V_\Lambda \,W^\Lambda +
V_{\Lambda\tau}\,W^{\Lambda\tau} - V^{\Lambda\tau} \,W_{\Lambda\tau} +\ft1{36}
\varepsilon^{\Lambda\Sigma\Gamma\Omega\Pi\Delta}
V_{\Lambda\Sigma\Gamma}\, W_{\Omega\Pi\Delta}   \,.
\ee
With this form of $\Omega^{MN}$ we can write down the constraint
\eqn{quadratic-constraint} for the embedding tensor defined by
\eqn{2-theta}, 
\be
\label{20-constraint}
\varepsilon^{\Lambda\Sigma\Gamma\Omega\Pi\Delta}
\,\theta_{\Lambda\Sigma\Gamma}{}^\tau\,
\theta_{\Omega\Pi\Delta}{}^\sigma  =0\;.
\ee
This constraint\footnote{
Note that from the perspective of dimensionally compactified IIB
theory, this constraint can be written as 
$\int_{T^6} G^\tau\wedge G^\sigma = 0$.} 
ensures that the gauge generators \eqn{2-theta} close under
commutation. Indeed, using \eqn{20-constraint} and \eqn{E7-positive}
we derive the gauge algebra, with the following non-vanishing
commutation relations,  
\bea
\label{gauge-algebra}
{[}X_{\Lambda},X_{\Sigma}]&=& - 2\,
\varepsilon_{\tau\sigma}\,\theta_{\Lambda\Sigma\Gamma}{}^\tau \,
X^{\Gamma \sigma} - \theta^\Gamma\,
X_{\Lambda\Sigma\Gamma}\;,\nonumber\\[1ex]
{[}X_{\Lambda},X^{\Sigma \tau}]&=& - \ft16
\varepsilon^{\Sigma\Gamma_1\Gamma_2\Gamma_3\Omega_1\Omega_2}\,
\theta_{\Gamma_1\Gamma_2\Gamma_3}{}^\tau\, X_{\Lambda\Omega_1\Omega_2}
\;. 
\eea
Clearly the gauge algebra is nilpotent. 
\paragraph{Some relevant \Exc7 transformations} 
Introducing the parameters $B_{\Lambda\Sigma\tau}$ and 
$\bar B^{\Lambda\Sigma\tau}$ for the generators 
with weight~$\pm 1$, $C^{\Lambda\Sigma}$ and $\bar C_{\Lambda\Sigma}$
for the generators with weight~$\pm 2$, and $D_\tau$ and $\bar D^\tau$
for the generators with weight~$\pm 3$, we can write down a
block-decomposition for an infinitesimal \Exc7 transformation in the
${\bf 56}$ representation. The
diagonal blocks are associated with ${\rm SL}(6)\times{\rm
SL}(2)\times {\rm SO}(1,1)$, which have weight~$0$. To define the
action of the nonzero-weight generators we list the
variation $\d V = \ft12(B_{\Lambda\Sigma \tau}\,t^{\Lambda\Sigma \tau}+ 
\bar B^{\Lambda\Sigma \tau} \,t_{\Lambda\Sigma \tau})V$ for the various
components,  
\bea
\label{B-Bbar}
\delta V^\Lambda              &=&  \bar B^{\Lambda\Sigma \tau}\, 
V_{\Sigma \tau}\;,     \nonumber\\
\delta V_{\Lambda \tau}          &=& B_{\Lambda\Sigma \tau}\,V^\Sigma +
\ft12  \varepsilon_{\tau\sigma}\, \bar B^{\Sigma\Gamma \sigma}
\,V_{\Lambda\Sigma\Gamma} \;,  \nonumber\\ 
\delta V_{\Lambda\Sigma\Gamma}&=&  3\,\varepsilon^{\tau\sigma}
\,B_{[\Lambda\Sigma \tau}\, V_{\Gamma]\sigma} -\ft1{2}
\varepsilon_{\tau\sigma} 
\,\varepsilon_{\Lambda\Sigma\Gamma\Omega\Pi\Delta} \,\bar B^{\Omega\Pi
\tau} \,V^{\Delta \sigma} \;,  \nonumber\\
\delta V^{\Lambda \tau}          &=& \ft1{12}  
\varepsilon^{\tau\sigma}\varepsilon^{\Lambda\Sigma\Gamma\Omega\Pi\Delta}\,
B_{\Sigma\Gamma \sigma} \, V_{\Omega\Pi\Delta} + \bar B^{\Lambda\Sigma
\tau} \,V_\Sigma\;,   \nonumber\\
\delta V_\Lambda             &=&  B_{\Lambda\Sigma \tau}\,V^{\Sigma \tau}
\;. 
\eea
The action of the remaining positive-weight generators follows from
imposing the \Exc7 commutators \eqn{E7-positive}. Observe that, up to
an overall normalization, \eqn{B-Bbar} is unique (up to an overall
normalization) upon assuming the 
condition $t^{\Lambda\Sigma \tau}= -(t_{\Lambda\Sigma
\tau})^\dagger$. From 
the commutation relations \eqn{E7-positive} one can then establish the
explicit form of the weight $\pm2$ and $\pm3$ generators through the
variation $\d V = \ft12 (C^{\Lambda\Sigma}\,t_{\Lambda\Sigma}+ 
\bar C_{\Lambda\Sigma } \,t^{\Lambda\Sigma})V+  (D_\tau t^\tau
+\bar{D}^\tau t_\tau)V$, 
which yields,
\bea
\delta V^\Lambda              &=&
\ft1{12}\varepsilon^{\Lambda\Sigma\Gamma\Omega\Pi\Delta}\, 
\bar C_{\Sigma\Gamma}\, V_{\Omega\Pi\Delta}
+ \varepsilon_{\tau\sigma}\bar D^\tau\,V^{\Lambda \sigma}\;,
\nonumber\\ 
\delta V_{\Lambda \tau}          &=& \varepsilon_{\tau\sigma}  \bar
C_{\Lambda\Sigma} \,V^{\Sigma \sigma} - \varepsilon_{\tau\sigma}\bar
D^\sigma \,V_\Lambda \;,  \nonumber\\ 
\delta V_{\Lambda\Sigma\Gamma}&=& 
\ft12 \varepsilon_{\Lambda\Sigma\Gamma\Omega\Pi\Delta} \, C^{\Omega\Pi}\,
V^\Delta +3 \,\bar C_{[\Lambda\Sigma}\, V_{\Gamma]} \;, \nonumber\\
\delta V^{\Lambda \tau}          &=&  
- \varepsilon^{\tau\sigma} \,C^{\Lambda\Sigma} \, V_{\Sigma \sigma}
+\varepsilon^{\tau\sigma}D_\sigma \,V^\Lambda\;, \nonumber\\
\delta V_\Lambda              &=&
-\ft12
C^{\Sigma\Gamma}\,V_{\Lambda\Sigma\Gamma}-\varepsilon^{\tau\sigma}D_\tau 
\,V_{\Lambda \sigma} \;. 
\eea
\paragraph{Construction of the $T$-tensor} 
To construct the $T$-tensor \eqn{T-theta}, we need the ${\rm
E}_{7(7)}/{\rm SU}(8)$ coset representative ${\cal V}$. A convenient 
decomposition is 
\be
\label{coset-representative}
{\cal V} = \exp[D_\tau\,t^\tau]\,\exp[\ft12
C^{\Lambda\Sigma}\,t_{\Lambda\Sigma}] \,\exp[\ft12
B_{\Lambda\Sigma \tau}\,t^{\Lambda\Sigma \tau}] \,\exp[\varphi\,t_0] \, \hat
{\cal V}(\Phi,\phi,\chi)\;,
\ee
where $\hat{\cal V}(\Phi,\phi,\chi)$ denotes the coset
representative of $[{\rm SL}(6)\!\times\!{\rm SL}(2)]/[{\rm
SO}(6)\!\times\!{\rm SO}(2)]$, which takes the form of direct
products of the coset representatives $L_\Lambda{}^I(\Phi)$ and  
$L_\tau{}^a(\phi,\chi)$, depending on 20 fields $\Phi$,
and the dilaton and the axion, $\phi$ and $\chi$, respectively. 
The remaining 48 fields are provided by
$\varphi$, $B$, $C$ and $D$. Suppressing  $\hat{\cal
V}(\Phi,\phi,\chi)$ leads
to a tensor~${\cal T}$, whose components transform under 
${\rm SL}(6)\times {\rm SL}(2)\times {\rm SO}(1,1)$, which is
converted into the standard 
${\rm SU}(8)$ $T$-tensor upon appropriate contractions with 
$\hat{\cal V}(\Phi,\phi,\chi)$. 
The ${\rm SL}(6)\times {\rm SL}(2)\times {\rm SO}(1,1)$ symmetry is
linearly realized on ${\cal T}$ and on the fields $B$, $C$ and $D$,
provided that we let $\theta_{\Lambda\Sigma\Gamma}{}^\tau$ and
$\theta^\Lambda$ transform as spurion fields. In this way we can 
identify the representations of the components of the ${\cal
T}$-tensor and verify that they belong to the admissible
representations listed in table~\ref{912-decomposition}. This feature
depends crucially on the coefficients adopted in \eqn{2-theta} and
confirms that the 
embedding tensor and the corresponding $T$-tensor transform in the
${\bf 912}$ representation. It is  straightforward to show that the
fields $C$ and $D$ cannot appear, because they lead to unwanted
representations with ${\rm SO}(1,1)$ weights that are too high,
whereas the fields $B$ can appear at most linearly.  

The results for the tensor ${\cal T}_M\vert^\a$ take the following form, 
\bea
\label{cT-tensor}
{\cal T}_{\Lambda\Sigma\Gamma}\vert_\tau &=& -2\,
\varepsilon_{\tau\sigma}\,  
\theta_{\Lambda\Sigma\Gamma}{}^\sigma\, {\rm e}^{-3\varphi} \;,
\nonumber\\[1mm]
{\cal T}^{\Lambda \tau}\vert^{\Sigma\Gamma}{}&=& \ft16
\varepsilon^{\Lambda\Sigma\Gamma\Omega\Pi\Delta}\,
\theta_{\Omega\Pi\Delta}{}^\tau \, {\rm e}^{- 3\varphi} \;,  
\nonumber\\
{\cal T}^{\Lambda \tau}\vert_\sigma&=& 
\,\d^\tau{}_\sigma\,\Big[ 
\theta^\Lambda + \ft16 \varepsilon^{\Lambda
\Sigma_1\Sigma_2\Sigma_3 \Sigma_4\Sigma_5}\, B_{\Sigma_1\Sigma_2 \rho}
\,\theta_{\Sigma_3\Sigma_4\Sigma_5}{}^\rho\Big] \,{\rm e}^{-4\varphi} \;,
\nonumber\\
{\cal T}_{\Lambda}\vert_{\Sigma\Gamma \tau}{}&=&
-\varepsilon_{\tau\sigma}\,\theta_{\Lambda\Sigma\Gamma}{}^\sigma \,
{\rm e}^{-3\varphi} \;, 
\nonumber\\[1mm]
{\cal T}_{\Lambda}\vert^{\Sigma\Gamma}&=& \delta_\Lambda{}^{\![\Sigma}\, 
\Big[
\theta^{\Gamma]} + \ft16 \varepsilon^{\Gamma] 
\Sigma_1\Sigma_2\Sigma_3 \Sigma_4\Sigma_5}\, B_{\Sigma_1\Sigma_2 \tau}
\, \theta_{\Sigma_3\Sigma_4\Sigma_5}{}^\tau\Big] \,{\rm e}^{-4\varphi}
\;. 
\eea
Note that the ${\cal T}$-tensor contains two independent
structures, associated with the $({\bf 20},{\bf 2})_{+3}$ and
$(\overline{\bf6},{\bf 1})_{+4}$ representations. This feature depends
crucially on the parametrization adopted for the coset representative
\eqn{coset-representative} with fields of non-negative weight. Because
of that the weights of the ${\cal T}$-tensor can only be equal to 3 or
4, which, in view of table~\ref{912-decomposition}, implies that only
two representations can be present. The ${\cal T}$-tensor should be gauge  
invariant. For a homogeneous space the isometries associated with the
gauge group can be calculated
straightforwardly. Choosing gauge group parameters
$\xi^{\Lambda\Sigma\Gamma}$, $\xi_{\Lambda \tau}$ and $\xi^\Lambda$,
these transformations read, 
\bea
\d B_{\Lambda\Sigma \tau}&=& -2\, 
\varepsilon_{\tau\sigma}\,\theta_{\Lambda\Sigma\Gamma}{}^\sigma \,
\xi^\Gamma\;, 
\nonumber\\[2mm]
\d C^{\Lambda\Sigma}&=& 2\,\xi^{[\Lambda} \theta^{\Sigma]}
-  \varepsilon^{\Lambda\Sigma\Gamma\Omega\Pi\Delta}
\Big( \ft13\theta_{\Gamma\Omega\Pi}{}^\tau\,\xi_{\Delta \tau} +\ft14
\theta_{\Gamma\Omega \Xi}{}^\tau\,B_{\Pi\Delta \tau}\,\xi^\Xi \Big) \;,
\nonumber\\
\d D_\tau &=& \varepsilon_{\tau\sigma}\,
\theta_{\Lambda\Sigma\Gamma}{}^\sigma \Big(-2\,
\xi^{\Lambda\Sigma\Gamma} + 
C^{\Lambda\Sigma}\,\xi^\Gamma \Big)  + 
\theta^\Lambda\, \xi_{\Lambda \tau}\nonumber\\
&& +  \ft1{24} \varepsilon^{\Lambda\Sigma\Gamma\Omega\Pi\Delta}
B_{\Lambda\Sigma \tau}\,B_{\Gamma\Omega \sigma}\,
\theta_{\Pi\Delta\Xi}{}^\sigma\,\xi^\Xi  \;,
\eea
which correctly generate the algebra \eqn{gauge-algebra}.
It is now straightforward to verify that the ${\cal T}$-tensor is gauge
invariant by virtue of the constraint \eqn{20-constraint}.

In order to obtain the ${\rm SU}(8)$ covariant $T$-tensors
$T_i{}^{j\,mn}$ and $T_{ijkl}{}^{mn}$ \cite{dWN82,dWST1}, which can be
decomposed into 
the tensors $A_1^{ij}$ and $A_{2\,i}{}^{\!jkl}$ that transform
according to the ${\bf 36}$ and ${\bf 420}$ representations (comprising
the ${\bf 912}$ representation of \Exc7), one must contract 
the previous ${\cal T}$-tensors with coset
representatives $L_\Lambda{}^I(\Phi)$ and $L_\tau{}^a(\phi,\chi)$
of ${\rm SL}(6)/{\rm SO}(6)$ and ${\rm SL}(2)/{\rm SO}(2)$,
respectively. The resulting tensors are assigned to representations of
${\rm SU}(4) \times {\rm U}(1)$ and take the form
\bea
\label{T-tensor}
{T}_{IJK}\vert_a &=& 
[L^{-1}(\Phi)]{}_I{}^\Lambda\, [L^{-1}(\Phi)]{}_J{}^\Sigma\,
[L^{-1}(\Phi)]{}_K{}^\Gamma\,
 [L^{-1}(\phi,\chi)]{}_a{}^\tau\; {\cal
T}_{\Lambda\Sigma\Gamma}\vert_\tau \;,
\nonumber\\[1mm]
{T}^{I a}\vert_b   &\equiv&
T^I\,\delta^a_b \nonumber\\
&=&  \d^a_b\, 
L_\Lambda{}^I(\Phi) 
\,\Big[ 
\theta^\Lambda + \ft16\varepsilon^{\Lambda
\Sigma_1\Sigma_2\Sigma_3 \Sigma_4\Sigma_5}\, B_{\Sigma_1\Sigma_2 \tau}
\,\theta_{\Sigma_3\Sigma_4\Sigma_5}{}^\tau\Big] \,{\rm e}^{-4\varphi} 
\;,
\eea
and likewise for all other components.  
From $T$, we build the (${\rm SO}(6)$ covariant) selfdual and 
anti-selfdual combinations,
\bea
{T}^+_{IJK\pm} &\equiv&
(T_{IJK}\vert_1\pm i\,T_{IJK}\vert_2) + 
\ft16{i}\,\varepsilon_{IJKLMN}\, (T_{LMN}\vert_1\pm
i\,T_{LMN}\vert_2)
\;,\nonumber\\[.5ex]
{T}^-_{IJK\pm} &\equiv&
\overline{{T}^+_{IJK\mp}}
\;.
\eea
The constraint (\ref{20-constraint}) takes the following form in terms
of the $T$-tensor, 
\bea
\label{T2-constraint}
T_{IJK+}^+\,T_{IJK-}^-\,=\,T_{IJK+}^-\,T_{IJK-}^+
\;. 
\eea
Once the embedding of ${\rm SU}(4) \times {\rm U}(1)$ 
into ${\rm SU}(8)$ is established, one can obtain the ${\rm SU}(8)$
covariant $T$-tensor. To 
see what the correct embedding is, we note that the generators of the
${\rm SU}(8)$ maximal compact subgroup of \Exc7  
consist of $t_{\Lambda\Sigma \tau} +t^{\Lambda\Sigma \tau}$, 
$t^{\Lambda\Sigma} +t_{\Lambda\Sigma}$, $t^{\tau} +t_{\tau}$, and those
belonging to  ${\rm SO}(6)\times{\rm SO}(2)$. Consequently, the
relevant embedding must take the form,
\begin{eqnarray}
{\bf 8} &\rightarrow & {\bf 4}_{+1/2} + {\bf 4}_{-1/2} \;, 
\nonumber\\ 
{\bf 63} &\rightarrow & {\bf 15}_0 + {\bf 15}_0 + {\bf 15}_{+1} + {\bf
15}_{-1} + {\bf 1}_0 + {\bf 1}_{+1} + {\bf 1}_{-1}\;.
\eea
Likewise we can give the corresponding branchings for the ${\bf
36}+{\bf 420}$ representation of the $T$-tensor. To work this out in
detail is somewhat laborious, but, as we shall show below, this
detailed information is not necessary for constructing the scalar
potential.   
\paragraph{Scalar potential}
The fermionic mass tensors and the scalar potential of the gauged
supergravity are usually given in terms of the tensors $A_1$ and
$A_2$. For the purpose of this discussion we want to remain in closer
contact with the $T$-tensor $T_M{}^\a$, which decomposes 
into two ${\rm SU}(8)$ tensors $T_i{}^{j\,mn}$ and $T_{ijkl}{}^{mn}$,
where the first indices $T_i{}^j$ and $T_{ijkl}$ refer to the compact
and noncompact \Exc7 generators generically denoted by the index
$\a$, while the index pair $[mn]$ refers to 28 of the indices $M$,
with the other 28 following from complex conjugation \cite{dWST1}. The
potential of gauged $N=8$ supergravity can now be written in the
following form, 
\bea
V &=& -\ft13 \vert A_1{}^{ij}\vert^2 + 
\ft1{24} \vert A_{2 i}{}^{jkl}\vert^2 
\nonumber \\
&=&- \ft 4{189} \Big[2\, \vert T_i{}^{jmn}\vert^2 -3\, \vert
T_{ijkl}{}^{mn}\vert^2 \Big] + \ft{17}{252} \vert
T_{ijkl}{}^{mn}\vert^2 \;.  
\eea
The first term is proportional to $\omega^{MN}\, T_M{}^\a
\,T_N{}^\b\,{\rm tr}(t_\a\,t_\b)$, where $\omega^{MN}$ is a symmetric
${\rm SU}(8)$ invariant tensor. The crucial point to note is that for
a nilpotent gauging such as \eqn{2-theta},
this terms vanishes. Therefore the potential is positive definite and
proportional to $\vert T_{ijkl}{}^{mn}\vert^2= \ft12 \vert A_2\vert^2$. 
In this way one deduces that it must take the  form, 
\bea
V(B,\varphi,\Phi,\phi,\chi) &=& 
\alpha\,T_{IJK+}^+(\varphi,\Phi,\phi,\chi)
\;T_{IJK-}^-(\varphi,\Phi,\phi,\chi) 
\nonumber\\[.5ex]
&&{}+ 
\beta\, T^I(\varphi,\Phi,B)\; T^I(\varphi,\Phi,B)
\;,
\label{potential}
\eea
with positive (group-theoretical) factors $\alpha$ and $\beta$ that
may be absorbed by 
rescalings of $\theta_{\Lambda\Sigma\Gamma}{}^\tau$, $\theta^\Lambda$ and
$B_{\Lambda\Sigma \tau}$.  

Let us briefly analyze the possibility of stationary points of this
potential. The form of (\ref{cT-tensor}), (\ref{potential}) shows that 
stationarity implies that $T^I$ must vanish at the stationary point, fixing
some of the  $B_{\Lambda\Sigma \tau}$ as functions
of~$\theta^\Lambda$. In view of the fact that 
the remaining part of $V$ depends exponentially on the field
$\varphi$, stationary points of $V$ can
only appear at $V=0$, requiring that $T_{IJK+}^+$ must vanish as
well. The constraint (\ref{T2-constraint}) then implies that also
$T^-_{IJK+}$ vanishes and therefore
$\theta_{\Lambda\Sigma\Gamma}{}^\tau=0$. In that case we are left with
the second term in the potential, which no longer leads to a
stationary point either. Therefore the potential \eqn{potential} has
no stationary 
points. 

Alternatively one may search for domain wall solutions, in which
only the field $\varphi$ is evolving while all other fields are
constant. The existence of such a solution
poses severe restrictions on the possible flux configurations. 
Here we present a simple example of a possible three-form
flux. To be specific, consider the index split $6\rightarrow3+3$ of
$\Lambda\rightarrow (\a,\hat\a)$, where both type of indices run from
1~to~3, with $\theta^\Lambda=0$ and 
$\theta_{\Lambda\Sigma\Gamma}{}^\tau$ taking the values 
\begin{eqnarray}
\label{special-flux}
&&\theta_{\a\b\g}{}^2 = -\kappa\,\theta_{\a\b\g}{}^1
= -\kappa\,b\,\varepsilon_{\a\b\g}
\;,\qquad
\theta_{\hat{\a}\b\g}{}^1 = -\kappa\,\theta_{\hat{\a}\b\g}{}^2 =  
\kappa \,b_{\hat \a} \,\varepsilon_{\hat{\a}\b\g}
\;, \nonumber \\[1ex]
&&
\theta_{\hat{\a}\hat{\b}\g}{}^2 =
-\kappa\,\theta_{\hat{\a}\hat{\b}\g}{}^1 = 
\kappa\, b_\g \,\varepsilon_{\hat{\a}\hat{\b}\g}
\;,\qquad
\theta_{\hat{\a}\hat{\b}\hat{\g}}{}^1 = 
-\kappa\,\theta_{\hat{\a}\hat{\b}\hat \g}{}^2 = 
\kappa\,b\,\varepsilon_{\hat{\a}\hat{\b}\hat \g}
\;,
\end{eqnarray}
parametrized by a constant vector $b_\a$ of length $b$ and a constant
$\kappa$. One can verify that this choice satisfies the constraint
\eqn{20-constraint}, and, in addition, 
\bea
\label{trace}
&&(1+\kappa^2)\Big[ 
\theta_{\Lambda\Gamma\Omega}{}^1\,\theta_{\Sigma\Gamma\Omega}{}^1 + 
\theta_{\Lambda\Gamma\Omega}{}^2\,\theta_{\Sigma\Gamma\Omega}{}^2\Big]
+2\,\kappa 
\Big[\theta_{\Lambda\Gamma\Omega}{}^1\,\theta_{\Sigma\Gamma\Omega}{}^2 +
\theta_{\Lambda\Gamma\Omega}{}^2\,\theta_{\Sigma\Gamma\Omega}{}^1
\Big]
\nonumber \\
&&{}
= 4\,b^2 (1-\kappa^2)^2\,\delta_{\Lambda\Sigma}\,.
\eea
Upon truncation to $\phi$, $\chi$ and $\varphi$, the
potential (\ref{potential}) takes the form 
\begin{eqnarray}
\label{potential2}
V &\propto& 
 {\rm e}^{-6\varphi}\Big[ 
(1\!+\!\kappa^2)\, {\rm e}^{-2\phi} +
(1\!+\!\kappa^2) (1+\chi^2){\rm e}^{2\phi}   
- 4\,\kappa\, \chi \Big] \;,
\end{eqnarray}
where the dilaton-axion complex is parametrized according to 
\bea
L_\tau{}^b(\phi,\chi) =\left(\begin{array}{ll}{\rm e}^\phi &
\chi\, {\rm e}^\phi \\ 0 &{\rm e}^{-\phi}\end{array}\right) \;. 
\eea
The potential \eqn{potential2} satisfies $\partial_\phi V =
\partial_\chi V =0$ at the following values of the dilaton and axion  
fields, 
\begin{eqnarray}
e^{2\phi} = \frac{1-\kappa^2}{1+\kappa^2}\;,\qquad
\chi = \frac{2\kappa}{1-\kappa^2} \;.
\label{dilax}
\end{eqnarray}
We have suppressed the fields $\Phi$ but for the above values of the
axion and dilaton fields, one can prove that also $\partial_\Phi V=0$
at $\Phi=0$, by virtue of \eqn{trace}. Obviously the fields $B$, $C$,
and $D$ parametrize flat directions; a more detailed analysis
shows that also a subset of the $\Phi$ will parametrize flat
directions associated with the subgroup of ${\rm SL}(6)$ that leaves
\eqn{special-flux} invariant. Hence for the choice \eqn{special-flux}
we have established the existence of a one-dimensional submanifold 
parametrized by the field $\varphi$ that is stationary with respect to
all the remaining fields. From this we infer the existence of a domain
wall solution,  in which $\varphi=\varphi(r)$ is running, while the
dilaton and the axion take  constant values.

In fact, we have established a continuous family of such solutions,
related by ${\rm SL}(6)\times {\rm SL}(2)$ transformations on the
embedding tensor \eqn{special-flux}. By a suitable ${\rm SL}(2)$
transformation we can, for instance, set the axion and dilaton fields
to zero. The corresponding domain wall solutions can
be elevated to ten-dimensional solutions with only non-trivial
three-form fluxes, and constant dilaton and axion fields. We refrain
from giving further details. 

\paragraph{Truncation to $N=4$ orientifold models}
In this last paragraph we describe the gauged $N=4$
supergravity models constructed in \cite{AFT} as consistent
truncations of the $N=8$ gauged 
supergravity of this paper. These models describe type-IIB theory
compactified on a $T^{p-3}\times T^{9-p}/\mathbb{Z}_2$ orientifold in
which the orientifold projection involves, besides the world-sheet parity
operation, the space-inversion ${\rm I}_{9-p}$ of the coordinates
of $T^{9-p}$; the latter is transverse to a set of
space--time filling $Dp$-branes (where $ p=3,\,5,\,7,\,9$; for
$p=7$ the involutiveness of the orientifold operation requires 
an additional action of $(-)^{F_L}$). 
The low-energy description of these orientifold compactifications
which involve some of the IIB fluxes, takes the form of a gauged,
four-dimensional $N=4$ supergravity~\cite{AFT}. Below we show how the
corresponding gauge algebras are 
obtained as suitable truncations of the $N=8$ gauge algebra
discussed in the present paper, once the fluxes in the various
$N=4$ models are identified with the components
$\theta_{\Lambda\Sigma\Gamma}{}^\tau$ and $\theta^\Lambda$ of the
embedding tensor.
 
Let us label the Neumann directions of
$T^{p-3}$ by $i,j,\dots=1,\dots, p-3$, and the Dirichlet directions
of $T^{9-p}$ on which ${\rm I}_{9-p}$ acts by $u,v,\ldots=p-2,\dots,
6$. The scalars $B_{\Lambda\Sigma}$ and $C_{\Lambda\Sigma}$
originating from the NS-NS and R-R two-forms are identified with the
fields $B_{\Lambda\Sigma \,1}$ and $B_{\Lambda\Sigma \,2}$ introduced
previously, while the four-form fields $C_{\Lambda\Sigma\Gamma\Omega}$
are related to the fields $C^{\Lambda\Sigma}$. Furthermore,
$G,\,H$ and $F$ denote the field strengths of the R-R and NS-NS
two-forms and of the R-R four-form,  
respectively.

Let us now discuss the models with $p=3,5,7$ (the case $p=9$ does not
allow nontrivial fluxes, and therefore there is no corresponding
gauging). The case $p=3$ corresponds to a IIB-compactification on a
$T^6/\mathbb{Z}_2$ orientifold. Apart from the ten-dimensional R-R
scalar $C_{0}$, the only non-metric axions that survive the projection
are the four-form internal components $C^{uv}\propto
\varepsilon^{uvwxyz}\, C_{wxyz}$ parametrizing the target-space
directions associated with the nilpotent generators $t_{uv}$. The
vector fields are $B_{u\mu}$ and $C_{u\mu}$, collectively denoted by
$A_{u \tau \mu}$, which couple to generators $X^{u \tau}$, while the
allowed fluxes are $G_{uvw}$ and $F_{uvw}$, denoted by
$\theta_{uvw}{}^\tau$. In this case the gauge group is abelian and
is generated by $X^{u \tau}\propto
\varepsilon^{uvwxyz}\,\theta_{vwx}{}^\tau\,t_{yz}$. These generators
clearly form an abelian subalgebra of the $N=8$ algebra
(\ref{gauge-algebra}) which does not involve the generators
$X_\Lambda$ and $X^{\Lambda\Sigma\Gamma}$. The closure of the $N=4$
algebra, $[X^{u\tau},\,X^{v\sigma}]=0$, does not require any condition
on the fluxes and therefore the most general $N=4$ model of this kind
can violate the quadratic constraint \eqn{20-constraint} which holds
on the fluxes in the $N=8$ model, namely $\int_{T^6} G\wedge H
\neq\,0$. Hence, the $T^6/\mathbb{Z}_2$ orientifold model can, in
general, not be embedded in the $N=8$ model.

The $p=5$ model corresponds to a compactification on a $T^{2}\times
T^{4}/\mathbb{Z}_2$ orientifold. The non-metric axions that survive
the projection  are 
$B_{iu},\,C_{ij},\,C_{uv},\,C_{iuvw}$ and $C_{\mu\nu}$ (dualized to a
scalar), which parametrize the target-space submanifolds associated
with the nilpotent generators  
$t^{iu\,1},\,t^{ij\,2},\,t^{uv\,2},\,t_{iu}$ and $t^1$,
respectively. The vector fields are 
$G^i_\mu,\,B_{u\,\mu},\,C_{i\,\mu}$ and $C_\mu{}^{\!uij}\propto
\varepsilon^{uvwr}\varepsilon^{ij} C_{vwr\,\mu}$ and couple to the
generators $X_i$, $X^{u1}$, $X^{i2}$ and $X_{uij}$. The
allowed fluxes are $H_{uvw},\,G_{iuv},\,H_{iju}=\varepsilon_{ij}\,H_u$
and $F_{ijuvw}$. For simplicity, let us switch on only $G_{iuv}$
and $H_u$. Identifying $\theta_{iju}{}^2$ with $-\varepsilon_{ij} H_{u}$ and
$\theta_{iuv}{}^1$ with $G_{iuv}$ and setting the remaining
$\theta$-components to zero we obtain for the generators of the gauge
algebra, using \eqn{2-theta},  
\begin{eqnarray}
X_\Lambda &\rightarrow & X_i\propto H_u\,\varepsilon_{ij}\,t^{ju
\,1}+G_{iuv}\,t^{uv\,2}\;, \nonumber\\
X^{\Lambda \tau} &\rightarrow &\cases{
X^{u1}\propto\varepsilon^{uvwr}\varepsilon^{ij}\,G_{ivw}\,t_{jr}\;,\cr 
X^{i2} =0\;,\cr}
\nonumber\\
X_{\Lambda \Sigma\Gamma} &\rightarrow &
X_{uij}\propto \varepsilon_{ij}\, H_u\,t^1 \;.
\end{eqnarray}
These generators span a subalgebra of the $N=8$ algebra
\eqn{gauge-algebra}, with nonvanishing commutators,
\begin{eqnarray}
[X_i,\,X_j]&\propto&\varepsilon_{ij}\,H_u
X^u\;, \qquad  [X_i,\,X^{u1}]\propto \varepsilon^{uvwr}\,\varepsilon^{jk}
\, G_{kvw}\,X_{ijr}\;, 
\end{eqnarray}
which coincides with the gauge algebra found in \cite{AFT}. The
knowledge of the full $N=8$ algebra \eqn{2-theta} makes it
straightforward to compute the extension  of the above $N=4$ algebra
in the presence of the remaining allowed fluxes, $H_{uvw}$ and
$G_{ijuvw}$. We note that the 
allowed fluxes in this model satisfy the quadratic
constraint $\int_{T^6}G \wedge H=0 $, so that this embedding is
consistent.  

The case $p=7$ corresponds to a compactification on a $T^{4}\times 
T^{2}/\mathbb{Z}_2$ orientifold  The non-metric axionic fields
surviving the orientifold projection, aside from $C_0$, are
$B_{iu},\,C_{iu},\,C^{uv}\sim\varepsilon^{ijk\ell uv
}\,C_{ijk\ell}$ and 
$C^{ij}\sim \varepsilon^{ijk\ell uv}C_{k\ell uv}$,
parametrizing the target-space directions associated with the
nilpotent generators $t^{iu \,1},\,t^{iu\,
2},\,t_{uv},\,t_{ij}$. The vector fields are $G^i_\mu$,
$B_{u\mu}$, $C_{u\mu}$ and $C_\mu{}^{\!iuv}\propto \varepsilon^{ijk\ell}
\varepsilon^{uv} C_{jk\ell\,\mu}$ which couple to the generators
$X_i$, $X^{u\tau}$ and $X_{iuv}$. The allowed fluxes are
$G_{iju}$, $H_{iju}$ and $F_{uvijk}$, which we identify with the
embedding tensor components according to $H_{iju}=- \theta_{iju}{}^2$, 
$G_{iju}=\theta_{iju\,1}$ and 
$\theta^i=\varepsilon^{ijk\ell}\varepsilon^{uv}\,F_{jk\ell uv}$, the
generators of the gauge algebra take the following form, where we
again use \eqn{2-theta}, 
\begin{eqnarray}
X_\Lambda &\rightarrow & 
X_i\propto H_{iju}\,t^{ju1} + G_{iju}\,t^{ju2}
+\varepsilon^{jk\ell m}\,F_{ijkuv}\,t_{\ell m}\;,
\nonumber\\ 
X^{u a} &\rightarrow &
\cases{X^{u1} \propto \varepsilon^{uv}\varepsilon^{ijk\ell}
\,G_{vij}\,t^{kl}\cr 
X^{u2} \propto  - \varepsilon^{uv}\varepsilon^{ijk\ell}\,
H_{vij}\,t^{kl}\;, \cr} \nonumber\\ 
X_{\Lambda\Sigma\Gamma} &\to& X_{iuv} =0\,.
\end{eqnarray}
They span a subalgebra of \eqn{gauge-algebra}, whose only non-trivial 
commutator is 
\begin{eqnarray}
[X_i,\,X_j]&\propto & H_{iju}\, X^{u1} + G_{iju}\, X^{u2} \;,
\end{eqnarray}
provided the fluxes satisfy the quadratic 
constraint ~\eqn{20-constraint}. Again this algebra is consistent with
the results obtained in \cite{AFT}.

\paragraph{Conclusions}
In this paper we demonstrated the use of the recently developed
group-theoretical approach for constructing gaugings of maximal
supergravity theories. The example that we presented is of relevance
for  compactifications of IIB theory with fluxes. In principle the
method is applicable to any 
gauging in any number of space-time dimensions. Furthermore we have
exhibited how consistent orientifold projections lead to a
variety of $N=4$ gauged supergravities. 
\vspace{8mm}

\noindent
{\bf Acknowledgement}\\
\noindent
This work is partly supported by EU contracts HPRN-CT-2000-00122
and HPRN-CT-2000-00131 and by the INTAS contract 99-1-590. M.T.\ was 
supported by a European Community Marie Curie Fellowship under
contract HPMF-CT-2001-01276. 

\bigskip
%

\end{document}